\documentclass[pra,aps,floatfix,twocolumn,showpacs,reprint,superscriptaddress]{revtex4}

\usepackage{bm}
\usepackage{epsfig}
\usepackage{mathptmx}
\usepackage{amssymb}
\usepackage{amsmath}
\usepackage{natbib}
\usepackage{color}
\usepackage[latin1]{inputenc}
\usepackage[T1]{fontenc}
\usepackage[loose,nice]{units}       

\newcommand{\Tr}{\operatorname{Tr}}
\newcommand{\ket}[1]{|#1\rangle}
\newcommand{\bra}[1]{\langle#1|}

\newcommand{\etal}{\textit{et al.}}

\begin{document}

%
%
\title{The role of final state correlation in double ionization of helium: a master equation approach}
\author{S{\o}lve Selst{\o}}
\affiliation{Centre of Mathematics for Applications, University of Oslo, N-0316 Oslo, Norway}
\author{Tore Birkeland}
\affiliation{Department of Mathematics, University of Bergen, N-5020 Bergen, Norway}
\author{Simen Kvaal}
\affiliation{Centre of Mathematics for Applications, University of Oslo, N-0316 Oslo, Norway}
\author{Raymond Nepstad}
\affiliation{Department of Physics and Technology, University of Bergen, N-5020 Bergen, Norway}
\author{Morten F{\o}rre}
\affiliation{Department of Physics and Technology, University of Bergen, N-5020 Bergen, Norway}

%
%
\date{\today}

%
%
\begin{abstract}
The process of nonsequential two-photon double ionization of helium is studied
by two complementary numerical approaches. First, the time-dependent
Schr{\"o}dinger equation is solved and the final wave function is analyzed in
terms of projection onto eigenstates of the uncorrelated Hamiltonian, i.e.,
with no electron-electron interaction included in the final states.  Then, the
double ionization probability is found by means of a recently developed approach in which
the concept of absorbing boundaries has been generalized to apply to systems
consisting of more than one particle. This generalization is achieved through
the Lindblad equation. 
A model of reduced dimensionality, which describes the process at a qualitative level, has been used. The agreement between the methods provides a strong indication that
procedures using projections onto uncorrelated continuum states are adequate when
extracting total cross sections for the direct double ionization process.
\end{abstract}

%
%











\maketitle

\section{Introduction}
The problem of multiphoton single- and multiple ionization of atoms and
molecules has been subject of intense research in recent years. The development
of new high-frequency light sources, such as free-electron lasers (FEL) and
high-order harmonic generation sources, capable of generating intense coherent
radiation, is one reason for this.  Pulses of attosecond durations are now
routinely produced by high-order harmonic generation with intensities as high
as 10$^{14}$ W/cm$^2$ \cite{mashiko}, enabling experimental studies of
nonlinear phenomena in the extreme-ultraviolet regime in
atoms~\cite{miyamoto,Nabekawa} and molecules~\cite{hoshina}.  Owing to recent
advances in FEL technology, femtosecond pulses of unprecedentedly high
intensity are now available, covering wavelengths ranging from vacuum
ultraviolet to soft x-rays~\cite{Ackermann, Shintake2008}.  These pulses have
opened the door for experimental studies of multiphoton multiple ionization of
complex atoms~\cite{Sorokin_PRL,Moshammer,Laarmann} and atom
clusters~\cite{Wabnitz_Nature}.  A parallel development of {\it ab initio}
numerical methods, capable of addressing the problem of intense-field
multiphoton ionization in one- and two-electron systems has taken
place~\cite{Parker_PRL_2006,Parker2001,Moore}.

Despite these developments, the problem of nonsequential (direct) two-photon
double ionization of helium still remains an unsolved problem, even though it
has been widely studied during the last decade both
theoretically~\cite{Pindzola,Feng2003, Bachau2003,
Bachau_EPD,Hu2005,Foumouo2006, Shakeshaft2007, Ivanov2007, Horner2007,
Nikolopoulos2007, Feist2008,Guan2008,Foumouo2008,Palacios2009,Nepstad2010}
and experimentally~\cite{Hasegawa_PRA_2005,Nabekawa, Sorokin2007,Rudenko}.  The
main obstacle in experiments has been to produce sufficiently high ionization
yields.  On
the theoretical side, accurate predictions for the total (generalized) cross
sections of the process remain elusive, as values obtained with different
methods vary by more than an order of magnitude.  What characterizes this
particular two-photon process is that it depends upon exchange of energy
between the outgoing electrons, i.e., it is a so-called nonsequential or direct
process, as opposed to a sequential ionization process where both electrons may
be considered as independent particles.  The sequential process is
energetically inaccessible for photon energies below 54.4 eV, but becomes the
dominant two-photon ionization process at higher energies.  Concerning the
nonsequential process, the great discrepancies that remain between different
theoretical calculations are usually ascribed to the different ways electron
correlations are handled.  It has furthermore been claimed, by several
authors, that the two-photon double ionization process in helium is a problem
in which electron correlations in the final state play an extremely important
role.  This stands somewhat in contrast to the related (direct) one-photon
double ionization process, where a complete agreement between different
theoretical approaches and experiments is now
established~\cite{Briggs,Huetz,Samson1998,Foumouo2006}.

\begin{figure}
	\begin{center}
		\includegraphics[width=8.5cm]{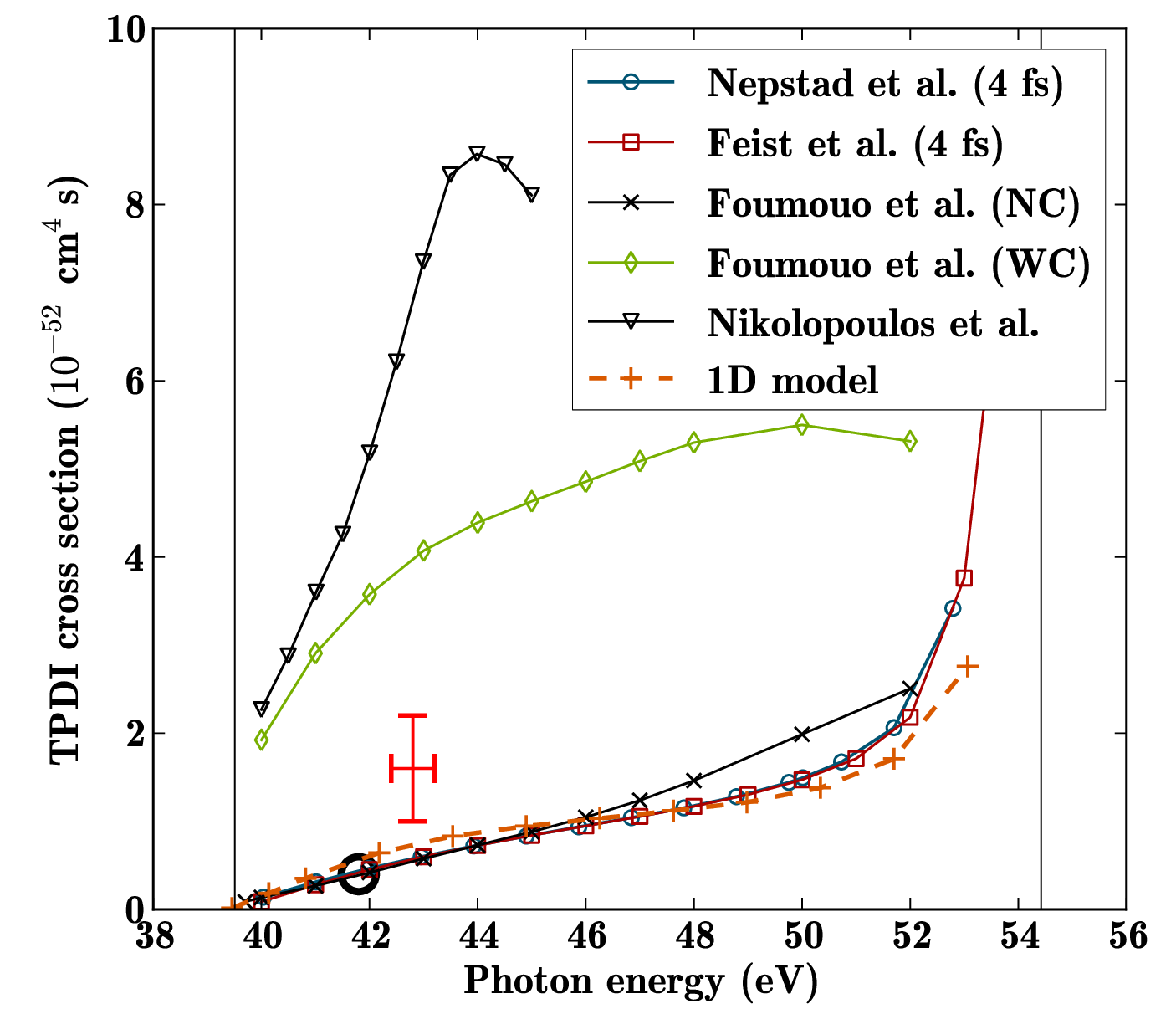}
	\end{center}
	\caption{(Color online) Two-photon double ionization cross sections.
	Black circle: experimental result of
	Hasegawa \etal~\cite{Hasegawa_PRA_2005}, red cross:
	experimental result of Sorokin
	\etal~\cite{Sorokin2007}, 
	blue line with circles: the results obtained by Nepstad \etal~\cite{Nepstad2010},
	red line with squares: Feist \etal~\cite{Feist2008}, green line
	with diamonds: Foumouo \etal~\cite{Foumouo2006} (with correlation, WC), black line with
	crosses: Foumouo \etal~\cite{Foumouo2006} (no correlation, NC), 
	orange dashed line with '$+$' (scaled): 1D result with the Hamiltonian~(\ref{Hamiltonian}),
	and black line with triangles:
	Nikolopoulos~\etal~\cite{Nikolopoulos2007}. The
	vertical lines define the two-photon direct double ionization region.
	}
	\label{DifferentSigmas}
\end{figure}

As it turns out, it is especially the separation of the two-photon single
ionization, where the remaining electron is left in an excited ionic state, from
the two-photon double ionization that is the bottleneck, the main problem being
that electrons are emitted with similar energies in the two processes.
Moreover, the fact that the single ionization process is much more probable
makes the problem even more challenging, and an exact knowledge of  the role of
correlation in the final state particularly important.  The effect of electron
correlations in the final state on the total integrated cross section was
studied in detail by Foumouo {\it et al.}~\cite{Foumouo2006,Foumouo2008}.  Using
the J-matrix method they were able to incorporate correlation effects in the
final (single continuum) scattering states, with the result that their
calculated generalized cross section increased significantly compared to the
result they obtained when projecting the final wave packet directly onto
uncorrelated products of Coulomb waves (with nuclear charge $Z=2$) (compare full
green curve with diamonds with full black curve with crosses in
Fig.~\ref{DifferentSigmas}). More importantly, their uncorrelated result is in
complete accordance with calculations performed by
others~\cite{Pindzola,Bachau2003, Bachau_EPD,Hu2005,Shakeshaft2007,
Feist2008,Guan2008,Nepstad2010}, which all have in common that the role of
correlation was completely neglected in the final state.  For comparison, these
results, together with the result  of Feist {\it  et al.}~\cite{Feist2008} and
Nepstad {\it et al.}~\cite{Nepstad2010}, and available experimental data,
is shown in Fig.~\ref{DifferentSigmas}.

%
%

Generating correlated multichannel states, and employing lowest-order
(non-vanishing) perturbation theory, Nikolopoulos and
Lambropoulos~\cite{Nikolopoulos2007} obtained a value for the cross section that
differed even more substantially from the uncorrelated results (black line with
triangles in Fig.~\ref{DifferentSigmas}). From this observation, it appears reasonable to conclude
that the inclusion of electron correlation in the final states is required.
However, the R-matrix calculations by Feng and van der Hart~\cite{Feng2003},
calculations employing exterior complex scaling by  Horner {\it et
al.}~\cite{Horner2007}, and the time-dependent close-coupling calculations by
Ivanov and Kheifets~\cite{Ivanov2007}, which all consider final state
correlations, produce results that are similar to the uncorrelated results.
Thus, at present, the matter remains unresolved.

In this article, we revisit the problem of nonsequential two-photon double
ionization of helium. We employ an approach based on the concept of
absorbing boundaries and the Lindblad formalism~\cite{Selsto2010} to calculate
the generalized cross section for the process.  Absorbing boundary conditions
are often quite useful in the context of wave packet propagation
\cite{Kulander1987,Fevens1999,Feuerstein2003}. 
However, applying them in the
study of systems consisting of several particles is not straightforward. 
This is due to the fact that when one particle is absorbed, the Schr{\"o}dinger equation 
does not retain any information about the remaining particles.
Here
we will demonstrate how the notion of absorbing boundary conditions may be generalized to apply to
many-particle systems in such a way that single and double ionization may be
distinguished~\cite{Selsto2010}. Specifically, the method allows for describing
the dynamics of the remaining electron as the first one is absorbed due to
photoionization.  A distinct advantage is that an exact knowledge of the final
scattering states is superfluous.  Furthermore, a direct comparison with the
result obtained by projecting the final wave function onto eigenstates of the
uncorrelated Hamiltonian reveals that this procedure indeed is adequate, as a
complete agreement between the two different methods is demonstrated.  

Describing a helium atom exposed to a laser pulse numerically is a challenging
task both in terms of  algorithm and computational power. Several theoretical
studies of this system involves models of reduced dimensionality
\cite{Lappas1998,Bauer1999,Lein2000,Haan2002,Zhang2010}. Only during recent
years has a full three-dimensional (3D) description become numerically
feasible. The formalism based on the Lindblad equation involves resolving the
dynamics of a one-particle density matrix in addition to a two-particle wave
function -- both involving six degrees of freedom when describing the full 3D
problem. Since these two sub-systems are coupled, more involved numerical
schemes are necessary when solving these equations than in the case of the
time-dependent Schr{\"o}dinger equation (TDSE). Thus, we resort to a
one-dimensional (1D) model atom in the present work. We argue, however, that
the coincidence between the double ionization probabilities obtained by the two
different methods do shed light on the question of whether projection onto
uncorrelated final states is an adequate approach also in the 3D case.

Having validated the Coulomb-wave projection method, we next consider the
convergence property of the cross section in the vicinity of the threshold at
54.4~eV.  It has been reported that the cross section exhibits an apparent
divergence in this limit~\cite{Horner2007,Shakeshaft2007,Horner2008,
Feist2008,Palacios2009}, a behavior that is usually interpreted as being due to
an unwanted inclusion of the sequential process in the calculations caused by
the bandwidth of the pulse~\cite{Lambropoulos_PRA_2008,Horner_2010}.  However,
it has recently been shown that this rise cannot solely be attributed to such
an effect~\cite{Nepstad2010}.
That 3D study was limited to photon energies below 52 eV. Going beyond the 52
eV limit would require the application of extremely long pulses with durations
of several tens of femtoseconds, which ultimately becomes an unmanageable
problem in 3D.  A 1D model remains computationally tractable, however, even for
these long pulses, thus allowing to study the behavior of the cross section very close
to the threshold. 

In the following section the theory is outlined. Specifically, Sec.~\ref{Model} describes the model and discusses to what extent it is able to make predictions also valid for the full 3D system. The two different methods used in order to obtain total cross sections are described in Secs.~\ref{Method1} and~\ref{Method2}. Their implementation, and possible extension to 3D is discussed in Sec.~\ref{Numerics}. The comparison between the two methods is presented in Sec.~\ref{ResAndDisc}. Furthermore, the behaviour of the cross section when the photon energy approaches the threshold at 54.4~eV is discussed. Conclusions are drawn in Sec.~\ref{Conclusion}.

\section{Theory}
\label{Theory}

%
%
\subsection{The model atom}
\label{Model}

The Hamiltonian of the system is given by
\begin{eqnarray}
\label{Hamiltonian}
H & = & h(x_1) + h(x_2) + V_\mathrm{Int}(x_1,x_2)\\
\label{SmallHamiltonian}
h(x) & = & -\frac{1}{2} \frac{d^2}{dx^2} + V_\mathrm{Nucl}(x) + x E(t)
\end{eqnarray}
where $x_i$ is the position of particle $i$ and $E$ is some external
time-dependent electric field of finite duration $T$.  Here, we have introduced atomic units,
denoted ``a.u.'', which are defined by choosing the electron mass, the electron
charge and $\hbar$ as units for their respective quantities. 
Both the interaction with
the nucleus, $V_\mathrm{Nucl}$, and the electron-electron interaction,
$V_\mathrm{Int}$, are taken as regularized Coulomb potentials, instead of bare
Coulomb potentials, i.e.,
\begin{eqnarray}
\label{NuclearPotential}
V_\mathrm{Nucl}(x) & = & -\frac{Z_\mathrm{Nucl}}{\sqrt{x^2+\delta^2}} \\
\label{ElecronInteraction}
V_\mathrm{Int}(x_1,x_2) & = & \frac{Z_\mathrm{Int}}{\sqrt{(x_1-x_2)^2+\delta^2}}.
\end{eqnarray}

By choosing $Z_\mathrm{Nucl} = 1.1225$, $Z_\mathrm{Int} = 0.6317 $ and $\delta =
0.3028$~a.u., the ground state and the first and second ionization thresholds
coincide with those of the 3D case. 
We emphasize that it is crucial that also the second ionization threshold, i.e., the first excited state of He$^+$, is correctly reproduced. 
For the photon energies
considered here, it means that one-photon excitations of He$^+$ is not possible,
concordant with the 3D case; indeed, such a possibility would introduce
artifacts detrimental to our present purpose.

While these are necessary conditions for the model to produce results comparable
with the original system, we cannot take for granted that
they allow for conclusions pertinent to the real 3D to be drawn.
Crucial geometrical aspects may potentially be lost when reducing the dimensionality of the problem. For instance, in the case of one-photon double ionization of helium, it is known that back-to-back scattering of the two electrons is prohibited for equal energy sharing~\cite{Foumouo2008}. Moreover, electron ejection is more probable in directions perpendicular to the polarization axis than parallel to it. Thus, for such a process, our 1D model cannot be expected to provide relevant information. 
For two-photon double ionization, however, it is known that back-to-back emission along the polarization axis is the dominating mode of ionization~\cite{Foumouo2008}. Such a process is indeed described within the 1D model. 

Predictions for the angular aspects of the double ionization process, such as the directional distribution of the photoelectrons, cannot be made within the model. However, when it comes to the total cross section, it has been argued that for the process at hand, radial correlations, as opposed to angular correlations, provide the dominant contribution~\cite{Piraux2008}. Hence, it seems reasonable to expect that the model should be able to describe the 
non-sequential double ionization process 
at a qualitative level. This assumption is strengthened by the fact that the 1D cross section, albeit too low in absolute value, does resemble the full 3D cross section as far as the photon energy is concerned, see Fig.~\ref{DifferentSigmas}.

In addressing the issue of whether projection onto uncorrelated final states provide the correct double ionization probability asymptotically, it is crucial not to underestimate the role of the electron-electron interaction. In the asymptotic region, both the nuclear potential and the electron-electron repulsion
of the model
coincide with the 3D case. Moreover, 1D models actually tend to {\it overestimate} correlation. This can be understood from the fact that the ``smoothing'' of the Coulomb potential, Eq.~(\ref{NuclearPotential}), effectively reduces the kinetic energy of the system, which in turn increases the importance of correlation~\cite{Birkeland2010}. In our model, this is confirmed by comparing the expectation value of the kinetic energy $T$ for the ground state with that of the original system; in 1D we find that $\langle T \rangle = 0.685$~a.u., which is about a fourth of the value found in 3D. The same tendency is found when comparing the correlation energy of the ground state; a Hartree-Fock calculation for the 1D model gives a ground state energy which is off by about 18~\% as opposed to 1.4~\% in the 3D case \cite{Wilson1935}. Thus, we expect electron-electron correlation to be at least as important in the 1D model as in 3D.

We note that, as the Hamiltonian is independent of electron spin, and
the initial state is a spin eigenstate with symmetric spatial part,
the system may be described formally as a two-boson system
\emph{without} spin. This will simplify the notation in the following.

%
%
\subsection{Uncorrelated final state approach}
\label{Method1}

In the method of projection onto uncorrelated final states, following Refs.~\cite{Pindzola,Bachau2003,
Bachau_EPD,Hu2005,Shakeshaft2007,
Feist2008,Guan2008,Nepstad2010}, we start out by diagonalizing the one-particle
Hamiltonian $h$ of Eq.~(\ref{SmallHamiltonian}), $h \, \phi_n = \epsilon_n
\phi_n$. The ``uncorrelated double continuum'' is defined by the span of
two-particle symmetric states with positive energies, 
\begin{eqnarray}
& \Phi^\mathrm{UCC}_{m,n} & =  N_{m,n} \left[\phi_m(x_1) \phi_n(x_2)+\phi_n(x_1) \phi_m(x_2) \right]
\label{UncorrelatedContinuum}
\\
\nonumber
& \mathrm{with} & \quad \epsilon_m,\epsilon_n  > 0 
\quad  \mathrm{and} \quad 
N_{m,n}=\left\{ \begin{array}{cl}\frac{1}{\sqrt{2}}, & m <n \\ \frac{1}{2}, & m=n  \end{array}\right. .
\end{eqnarray}
Clearly, these states do not represent the actual double continuum states of the
system.  However, assuming that the electron correlation diminishes in
significance as the doubly ionized wave packet travels outwards, the correct
double ionization probability should be obtained by projecting the unbound part
of the final state wave function at sufficiently long time after the interaction
onto the uncorrelated continuum states,
\begin{equation}
\label{DoubleIonizationUCC}
P_\mathrm{DI} = \sum_{m} \sum_{n \geq m} \left| \langle \Phi_{m,n}^\mathrm{UCC} | \mathcal{P}_\mathrm{UB} | \Psi(t \rightarrow \infty) \rangle \right|^2,
\end{equation}
where the projection operator $\mathcal{P}_\mathrm{UB}$ removes the bound part.

%
%
\subsection{Absorbing boundaries and the Lindblad equation}
\label{Method2}

We seek to determine the validity of the above approach by comparing the double
ionization yields thus obtained to the ones obtained using absorbing boundaries.
These are introduced by augmenting the Hamiltonian with a complex absorbing potential, 
$-i \Gamma(x)$.
The function $\Gamma(x)$ is zero within a certain region of $x$ and beyond this region 
it is positive and increasing with distance,
\begin{eqnarray}
\label{GammaFunk}
\Gamma(x) & =  & 0, \quad |x|\leq x_T \\
\nonumber
\Gamma(x) & > & 0, \quad |x| > x_T.
\end{eqnarray}
Waves entering into the region where $\Gamma>0$ are attenuated and eventually die out completely.

Simply introducing the absorber in the TDSE 
is problematic when we wish to
distinguish between single and double ionization. This becomes evident when
considering the fact that the wave function is normalized to the probability of
{\it all} particles being represented.  If one electron travels outwards and is
subsequently absorbed, not only is all information about this electron lost, but
so is all information about the remaining electrons as well. Thus, only
information about total ionization probabilities may be retained when using an
absorber in combination with the TDSE, while distinguishing between
single, double etc. ionization is hard to achieve. 

In a recent paper it was demonstrated how the notion of absorbing boundaries may
be generalized to an $N$-particle context in such a way that the remaining
$(N-1)$-particle system is recovered  after one particle has been absorbed from
the system~\cite{Selsto2010}.  Likewise, the $(N-2)$-particle system is
recovered when yet another particle is absorbed etc. It was argued that, since
this is a Markovian process, the Lindblad equation is the proper framework
within which to achieve this~\cite{Gorini1976,Lindblad1976}. 
In general, the dynamics of a system where particles are lost due to a complex absorbing potential is described by~\cite{Selsto2010}
\begin{eqnarray}
i \hbar \frac{d}{d t} \rho_n &=& [\hat{H}, \rho_n] - i \{ \hat{\Gamma}
,\rho_n \} +\notag \\ & &
2 i \int \Gamma(\xi) \boldsymbol{\psi}(\xi) \rho_{n+1} \boldsymbol{\psi}^\dagger(\xi) \, d\xi,
\label{LindbladGeneral}
\end{eqnarray}
where $\rho_n$ is the density operator corresponding to the $n$-particle sub-system, $n=0,1,...,N$. The generalized coordinates $\xi$ correspond to {\it all} degrees of freedom.
The operators $\boldsymbol{\psi}(\xi)$ and $\boldsymbol{\psi}^\dagger(\xi)$, which annihilate and
create, respectively, a particle with position and spin given by $\xi$, obey the usual
anti-commutation relations for fermions:
\begin{equation}
\label{AntiCommutator}
\{\boldsymbol{\psi}(\xi), \boldsymbol{\psi}(\xi') \} = 0, \quad \{\boldsymbol{\psi}(\xi), \boldsymbol{\psi}^\dagger(\xi') \}= \delta(\xi-\xi').
\end{equation}
The operators $\hat{H}$ and $\hat{\Gamma}$ are the Hamiltonian and the
complex absorbing potential, respectively, expressed in terms of these
field operators such that their explicit forms do not depend on the particular
number of particles. 
Specifically, for
the two-particle system at hand the
evolution may be written as
\begin{eqnarray}
\label{TwoPartLindblad}
i \hbar \frac{d}{d t} \ket{\Psi_2} & = & ( H - i \Gamma(x_1) - i \Gamma(x_2) ) \, \ket{\Psi_2}
\\
\label{OnePartLindblad}
i \hbar \frac{d}{d t} \rho_1 & = & [h,\rho_1] - i \{\Gamma,\rho_1\} +  
\\
\nonumber
& &
2 i \int \Gamma(x) \boldsymbol{\chi}(x) \ket{\Psi_2} \bra{\Psi_2} \boldsymbol{\chi}^\dagger(x) \, dx \\
\label{ZeroPartLindblad}
\hbar \frac{d}{dt} \rho_0 & = &  2 \int \Gamma(x) \boldsymbol{\chi}(x) \rho_1 \boldsymbol{\chi}^\dagger(x) \, dx,
\end{eqnarray}
where $\ket{\Psi_2}$ is the two-particle state, $\rho_1$ is a one-particle
density operator and $\rho_0=p_0(t) \ket{-} \bra{-}$ is a density operator
corresponding to vacuum, i.e., no particles present. In Eqs.~(\ref{TwoPartLindblad},\ref{OnePartLindblad},\ref{ZeroPartLindblad}) the spin degree of freedom is integrated out. Hence the field operators $\boldsymbol{\chi}^{(\dagger)}(x)$, which annihilate (create) a particle in position $x$, correspond formally to spinless bosons.
We emphasize that this formalism does not rely on the reduced dimensionality of the model. The equations remain the same regardless of the dimensionality of the coordinates $x$ (or $\xi$).

\begin{figure}[t]
	\begin{center}
		\includegraphics[width=8.5cm]{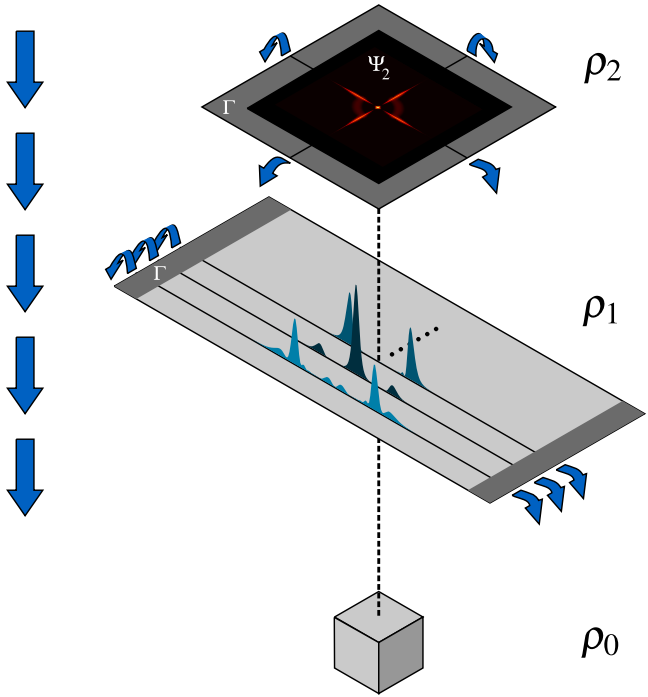}
	\end{center}
	\caption{(Color online) A schematic view of the absorbing boundary multi-particle method. The 
	system is initially described by a two-particle wave function (top). As a particle 
	is ionized and subsequently hits the absorber near the boundary ($\Gamma$), the remaining
	electron is described by a one-particle
	density matrix $\rho_1$ (middle level). Some of the eigenstates of 
	$\rho_1$ are shown. Further ionization cause also $\rho_1$ to overlap with the absorber, 
	which in turn 
	causes the vacuum state to be populated.
	See text for further details.}
	\label{fig:method_illustration}
\end{figure}

We see that $\Psi_2$ follows the TDSE.
Equation~(\ref{OnePartLindblad}), which governs the evolution of the one-particle sub-system,
has several terms. The first term on the right hand side, which corresponds to
the von Neumann equation, simply governs the evolution of $\rho_1$ dictated by
the one-particle Hamiltonian $h$. The second term corresponds to removal of the
particle due to the absorber, whereas the last term is a source term depending
on the overlap between the absorber and $\Psi_2$. The decrease in norm of $\Psi_2$ is
identical to the increase in the trace of $\rho_1$ induced by this source term,
and the remaining electron is properly reconstructed. Similarly, the
vacuum-probability $p_0$ increases by the same amount as $\Tr(\rho_1)$
decreases due to absorption. 
These mechanisms are illustrated schematically
in Fig.~\ref{fig:method_illustration}.
The sum of all norms and traces remains 
equal to unity at all times~\cite{Lindblad1976}, i.e.,
\begin{eqnarray}
\label{ConserveTrace}
& &p_2(t) + p_1(t) + p_0(t) = 1, \quad \forall t\\
\nonumber
& \mathrm{with} \ \ & p_2(t) \equiv \| \Psi_2(t) \|^2 \quad \mathrm{and} \quad p_1(t) \equiv \Tr(\rho_1(t)).
\end{eqnarray}

As also illustrated in Fig.~\ref{fig:method_illustration}, the density operator $\rho_1$ describes 
in general a mixed state,
\begin{equation}
\label{MixedState}
\rho_1(t) = \sum_n p_1^{(n)} \left| \Psi_1^{(n)}(t) \right\rangle \left\langle \Psi_1^{(n)}(t) \right|,
\end{equation}
i.e., one cannot in general expect a single-particle \emph{wave function}
since correlation information is lost whenever particles are removed. 

%
%

%
%
\begin{figure}
\begin{center}
\epsfxsize=8.5cm \epsfbox{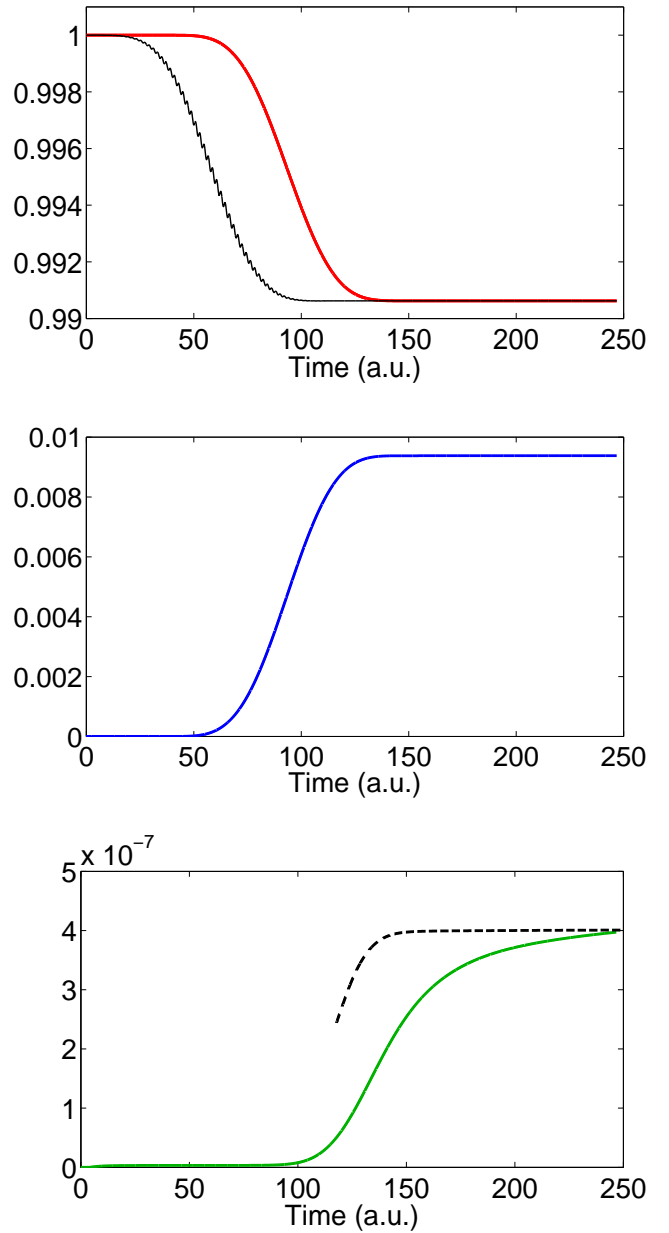}
\end{center}
\caption{(Color online) The probability of having two (upper panel), one (middle
panel) or zero particles (lower panel) on the grid. The thin, black curve in the
upper panel shows the probability for the system to remain in the ground state.
The pulse duration is here $T=$~118~a.u.~$= 2.86$~fs, the peak intensity is $2.2\times
10^{13}$W/cm$^2$, and the central frequency of the pulse corresponds to a photon
energy of 43.5~eV. We see that both the one-electron probability,
$p_1(t)=\Tr(\rho_1(t))$, and the zero-particle probability $p_0(t)$ increase as
the norm of the two-particle probability, $p_2(t)= \| \Psi_2(t)\|^2$, decreases.
The dashed curve in the lower panel represents the double ionization probability
obtained by analyzing the absorbed part of the wave function ``on the fly''. See
text for details.} 
\label{TimeConvergence}
\end{figure}

In Fig.~\ref{TimeConvergence} we have shown $p_2(t)$, $p_1(t)$ and $p_0(t)$ as
functions of time both during and after the interaction with the laser pulse. It
is seen that the two-particle probability decreases during and after the pulse,
whereas the one- and zero-particle probabilities increase -- all towards some
asymptotic value. It is also seen that $p_0(t)$ converges much more slowly than
$p_2(t)$ and $p_1(t)$. We will return to this issue later. In the upper panel
the probability for the system to remain in the ground state is also displayed.
It shows that practically all population of bound states is confined to the
ground state; very little excitation is seen.

The key point here is that as the first electron is absorbed, the dynamics of
the remaining one may still be described. Moreover, if also the second electron
is absorbed, we will eventually find the total system in the vacuum state.  Now,
suppose we may choose the absorption-free region large enough for 
the probabilities $p_0$, $p_1$ and $p_2$ to converge 
towards values independent of $x_T$, then these probabilities are
subject to straightforward interpretations: $p_2(t \rightarrow \infty)$ is the
probability that the system remains bound after interaction, $p_1(t \rightarrow
\infty)$ is the single ionization probability, and $p_0(t \rightarrow \infty)$
is the double ionization probability.  This interpretation relies on the idea
that there is a finite distance from the nucleus beyond which we may conclude
that the electron is indeed in the continuum. The validity of this assumption, in turn,
rests on whether the probabilities $p_1(t \rightarrow \infty)$ and $p_0(t
\rightarrow \infty)$ converge as the absorber is moved outwards.

The ideal absorber should be both transmission and reflection free, and hence
not influence the dynamics in the region where $\Gamma(x)=0$ at all. In a many-body context, a certain
indirect influence is difficult to avoid, however, since removal of one particle
also removes the interaction between this particle and the remaining ones. This may be crucial if the field moves the electrons such that they are close while one of them is being absorbed.
However, given a large enough $x_T$ and long enough propagation time, 
the electron-electron repulsion itself should cause the importance of this interaction to diminish. 
Hence we conclude, once again, that the approach should be valid if one is able to obtain absorber-independent results.

As in the case of projection onto uncorrelated final states, the state of the
system must be propagated a certain time beyond the duration $T$ of the
electric pulse in order for the double ionization probability $p_0$ to reach
convergence. Having obtained the probability of direct double ionization,
$P_\mathrm{DI}$, for a situation in which perturbation theory applies, the
total cross section for direct double ionization is found as~\cite{Feist2008,
Madsen2000}
\begin{equation}
\label{Sigma}
\sigma = \left( \frac{\omega}{I_0} \right)^2 \frac{P_\mathrm{DI}}{T_\mathrm{eff}}, \quad T_\mathrm{eff} \equiv \int_0^T \left( \frac{I(t)}{I_0} \right)^2 \, dt,
\end{equation}
where $I(t)$ is the intensity of the laser pulse, and $I_0$ is the maximum
intensity. For a pulse with a sine square envelope,
\begin{equation}
\label{SineSquare}
E(t) = E_0 \sin^2 \left( \frac{\pi t}{T} \right) \sin(\omega t + \varphi),
\end{equation}
we have $T_\mathrm{eff}=35 T/128$.

%
%
\subsection{Numerical considerations}
\label{Numerics}

As mentioned in the introduction, one of the reasons for choosing a 1D model atom rather than trying to solve the full 3D problem is related to the implementation of Eqs.~(\ref{TwoPartLindblad},\ref{OnePartLindblad},\ref{ZeroPartLindblad}). 
While there are no {\it formal} obstacles prohibiting the application of the
method to a 3D situation, and in spite of the fact that a grid of relatively
small extension may be used, such calculations would generally involve very
large basis sets. Thus, in a practical situation a parallelized version of the
scheme is required.
Since the two-particle dynamics is simply dictated by the Schr{\"o}dinger
equation, this part may be
handled by any of a number of proposed parallel numerical schemes available in the
literature~\cite{Smyth1981, Feist2008, Nepstad2010}, given the availability of
sufficient computational resources.
A similar scheme could then be adapted for the one-particle density matrix,
Eq.~(\ref{OnePartLindblad}). 
The remaining bottleneck is the calculation of the source term, which must be updated
at each time step. Every part of the two-particle wavefunction at the previous time step which has a finite overlap with the absorber will contribute to the source term.
This operation is potentially quite demanding on communication between the processing elements (PE), and some scheme to partially assemble the source term with the local data on each node is probably required, before these sub-elements may be transmitted to the correct PE for final assembly.
Alternatively, the complexity of solving Eq.~(\ref{OnePartLindblad}) could be
reduced by a lower rank approximation, e.g., along the lines proposed by Leth and
coworkers~\cite{Leth2009}.
In any case, a non-parallel 1D model implementation is a natural starting point
for further developments of the method, producing benchmarks and providing a
convenient testing ground for determining optimal implementations of the various
parts of the algorithm.

Returning to our 1D model, both the TDSE and
Eqs.~(\ref{TwoPartLindblad},\ref{OnePartLindblad},\ref{ZeroPartLindblad}) are
solved by means of split operator techniques~\cite{Feit1982} on uniform
numerical grids, $x_i^j=-L/2 +(j-1) \Delta x$, where $i=1,2$ is the particle
index, and $\Delta x=L/(\mathcal{N}-1)$, where $\mathcal{N}$ is the number of grid points. The code
solving the TDSE, implemented in the Pyprop
framework~\cite{pyprop, Nepstad2010}, is parallelized and thus able to
handle relatively large grids.

The numerical scheme applied in order to solve
Eqs.~(\ref{TwoPartLindblad},\ref{OnePartLindblad},\ref{ZeroPartLindblad}) was
presented in~\cite{Selsto2010}.  However, here we have made some adjustments to
the method in order to accommodate for two challenges which are both related to
the long-range nature of the Coulomb interaction with the nucleus.
Firstly, the system features Rydberg states, which may have a finite overlap
with the absorber. If these states are populated, it could induce
artificial ionization on large time scales. Hence, at $t=T$ all components of
the bound states of the helium system was removed. These states, including the
initial ground state, were found by propagation in imaginary time.  Secondly, it
is well known that photoelectrons of low kinetic energy may emerge from the
process at hand~\cite{Foumouo2008}. This may be somewhat problematic since, as
we have seen in Fig.~\ref{TimeConvergence}, it causes $p_0(t)$ to converge
rather slowly in time (it takes the second electron a long time to reach the absorber).
Moreover, the low energy of the photoelectron makes it vulnerable to artificial
reflections induced by the absorbing potential. After the interaction with the
laser pulse, $t>T$, this problem may be circumvented by analyzing the source
term in Eq.~(\ref{OnePartLindblad}) ``on the fly'' instead of propagating
$\rho_1$ further.  Suppose that the ``amount'' of double ionization is known at
a time $t>T$,  then at the next time step, $t+\Delta t$, the source term in
Eq.~(\ref{OnePartLindblad}) adds a contribution $(\Delta \rho_1)_S$ to the
one-particle density operator.  Since it is not very costly to diagonalize the
one-particle Hamiltonian $h$, as compared to diagonalizing the full two-particle Hamiltonian
$H$, this contribution $(\Delta \rho_1)_S$ may be separated into a part corresponding to bound
He$^+$ and a (double) continuum part. The double ionization probability is then
obtained simply by repeatedly adding (integrating) the latter contribution to
the double ionization probability. Consequently, it is not necessary to solve
Eqs.~(\ref{OnePartLindblad},\ref{ZeroPartLindblad}) for $t>T$ -- as long as the
absorbed part of $\Psi_2$ is properly analyzed. In the lower panel of
Fig.~\ref{TimeConvergence} it is clearly demonstrated that the converged value for
the double ionization probability is more easily obtained this way. 

Note that during
the interaction with the laser pulse, $t<T$, it \textit{is} necessary to resolve 
the one- and zero-particle dynamics as well, since the remaining electron is 
still subjected to the electric field. We would like to point out that the formalism 
based on the Lindblad equation allows for analysis of the ionization channel 
yields using a grid that is smaller than the actual extension of 
the {\it complete} ionized wave packet. This holds also during interaction with the laser pulse. 

For the Lindblad-formalism, practically convergent results for the double ionization 
probabilities were obtained using a box of 
size $L=$~130~a.u., a grid spacing of $\Delta x = 0.254$~a.u. and a 
numerical time step $\Delta t=2.50 \times 10^{-3}$~a.u. 
For the complex absorbing potential we have chosen a Manolopoulos form, c.f. \cite{Manolopoulos2002}, which has several advantages such as being completely transmission free and only containing one parameter. We have fixed this parameter by the choosing $x_T = 0.2 \,L$, c.f. Eq.~(\ref{GammaFunk}). 
Finally we note that, due to trace conservation, Eq.~(\ref{ConserveTrace}), it is not strictly necessary 
to solve Eq.~(\ref{ZeroPartLindblad}) along with Eqs.~(\ref{TwoPartLindblad},\ref{OnePartLindblad}). 
However, since our numerical scheme is not manifestly unitary, we have 
still done so in order to check that Eq.~(\ref{ConserveTrace}) indeed is fulfilled to a satisfactory degree.

In the TDSE calculations, where the final wave packet is projected 
directly onto uncorrelated products of Coulomb waves (with nuclear charge $Z=2$), Eq.~(\ref{DoubleIonizationUCC}), 
much larger boxes are required in order to contain the entire ionized
wavepacket during the time evolution. We used $L \leq \unit[2400]{a.u.}$,
which was sufficient for pulse durations up to $\unit[20]{fs}$, and photon
energies up to 53 eV. The grid resolution was fixed at $\Delta x =
\unit[0.195]{a.u.}$, while the time step was $\Delta t = 5.00 \times 10^{-3}$~a.u.

\section{Results and discussion}
\label{ResAndDisc}

%
%
\begin{figure}
\begin{center}
\epsfxsize=8.5cm \epsfbox{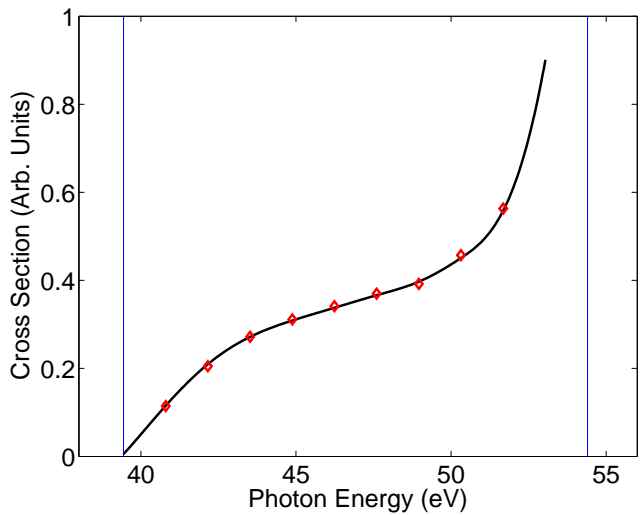}
\end{center}
\caption{(Color online) The (generalized) cross section for nonsequential
two-photon double ionization of our 1D model helium atom. The full curve
is obtained by solving the TDSE on a large grid with
subsequent projection of the final wave packet directly onto uncorrelated
products of Coulomb waves (with nuclear charge $Z=2$).  The  diamonds are
calculated by the method involving absorbing boundaries and the Lindblad
equation.}
\label{Coincidence}
\end{figure}

Figure~\ref{Coincidence} shows the total cross section for the process of
nonsequential double ionization by two photons. Results have been obtained using
both methods discussed above. In both cases, the electric field was given by
Eq.~(\ref{SineSquare}). For lower photon energies the duration of the field
corresponded to 30 optical cycles, whereas increasingly longer pulses where
necessary in order to obtain converged cross sections as the photon energy
increases towards the threshold at $\hbar \omega = 54.4$~eV. We see that the
cross sections obtained in the two different ways are in very good agreement.
This provides support for the proposition that the uncorrelated final states,
Eq.~(\ref{UncorrelatedContinuum}), are indeed able to provide accurate
information about photoelectrons emerging from helium asymptotically.

In~\cite{Feist2008} the strong dependence of the cross section as defined in
Eq.~(\ref{Sigma}) on the pulse length $T$ as $\hbar \omega \rightarrow 54.4$~eV
was discussed.  It was demonstrated that $\sigma$, which seems to have a sharp
rise, depends rather strongly on the pulse duration $T$ in this limit.  A
possible explanation for this was considered: for finite $T$, there is also a
finite spectral width of the pulse, enabling sequential double ionization also
for central frequencies slightly below threshold. As the probability of
sequential double ionization scales with $T^2$, as opposed to $T^1$ in the
nonsequential case, this gives rise to a pronounced increase in $\sigma$. The 4
fs pulse used in~\cite{Feist2008}, although not long enough to resolve the cross
section behavior very close to the threshold, was sufficient to observe the
start of an increasing trend, and thus rejects the sequential overlap hypothesis
with some confidence.
However, in order to resolve the true behavior of the cross section in the
immediate vicinity of the threshold, very narrow bandwidths, i.e., very long
pulses, are required.
By performing
calculations with increasing pulse duration (up to 20 fs), we have confirmed
that the value of $\sigma$ {\it does} converge towards a well-defined value as
$T$ increases -- also near the threshold. Moreover, as the spectral width of the
pulse becomes so narrow that the overlap with the region in which sequential
two-photon double ionization may take place vanishes, the pronounced rise in the
cross section is seen to prevail. Hence, the fact that $\sigma$ increases
sharply as $\hbar \omega$ approaches 54.4~eV cannot be due to the final
bandwidth of the pulse alone.

\section{Conclusion}
\label{Conclusion}

We have investigated the role of electron correlations in obtaining the cross
section for the process of nonsequential two-photon double ionization of helium,
using two different methods of approach -- by projection onto uncorrelated
continuum states, and by imposing a complex absorbing potential within the proper many-body framework. The former has
the advantage that these states are much easier to construct than the true
two-electron continuum states, whereas no such states are needed in the latter
method. Moreover, within this method single and double ionization yields may be 
obtained using a numerical grid considerably smaller than what is necessary to 
contain the ionized part of the  two-particle wave function. 
The predictions of the methods were in agreement, thus suggesting that projecting 
the ionized part of the full wave function onto uncorrelated continuum states does 
provide correct ionization yields in the asymptotic limit.

Finally, we presented evidence that the cross section for this process remains
well defined for photon energies arbitrarily close to the threshold at $\hbar
\omega= 54.4$~eV (from below). The cross section is seen to have a sharp
increase in this limit.

%
%


\begin{thebibliography}{57}
\expandafter\ifx\csname natexlab\endcsname\relax\def\natexlab#1{#1}\fi
\expandafter\ifx\csname bibnamefont\endcsname\relax
  \def\bibnamefont#1{#1}\fi
\expandafter\ifx\csname bibfnamefont\endcsname\relax
  \def\bibfnamefont#1{#1}\fi
\expandafter\ifx\csname citenamefont\endcsname\relax
  \def\citenamefont#1{#1}\fi
\expandafter\ifx\csname url\endcsname\relax
  \def\url#1{\texttt{#1}}\fi
\expandafter\ifx\csname urlprefix\endcsname\relax\def\urlprefix{URL }\fi
\providecommand{\bibinfo}[2]{#2}
\providecommand{\eprint}[2][]{\url{#2}}

\bibitem[{\citenamefont{Mashiko et~al.}(2004)\citenamefont{Mashiko, Suda, and
  Midorikawa}}]{mashiko}
\bibinfo{author}{\bibfnamefont{H.}~\bibnamefont{Mashiko}},
  \bibinfo{author}{\bibfnamefont{A.}~\bibnamefont{Suda}}, \bibnamefont{and}
  \bibinfo{author}{\bibfnamefont{K.}~\bibnamefont{Midorikawa}},
  \bibinfo{journal}{Opt. Lett.} \textbf{\bibinfo{volume}{29}},
  \bibinfo{pages}{1927} (\bibinfo{year}{2004}).

\bibitem[{\citenamefont{Miyamoto et~al.}(2004)\citenamefont{Miyamoto, Kamei,
  Yoshitomi, Kanai, Sekikawa, Nakajima, and Watanabe}}]{miyamoto}
\bibinfo{author}{\bibfnamefont{N.}~\bibnamefont{Miyamoto}},
  \bibinfo{author}{\bibfnamefont{M.}~\bibnamefont{Kamei}},
  \bibinfo{author}{\bibfnamefont{D.}~\bibnamefont{Yoshitomi}},
  \bibinfo{author}{\bibfnamefont{T.}~\bibnamefont{Kanai}},
  \bibinfo{author}{\bibfnamefont{T.}~\bibnamefont{Sekikawa}},
  \bibinfo{author}{\bibfnamefont{T.}~\bibnamefont{Nakajima}}, \bibnamefont{and}
  \bibinfo{author}{\bibfnamefont{S.}~\bibnamefont{Watanabe}},
  \bibinfo{journal}{Phys. Rev. Lett.} \textbf{\bibinfo{volume}{93}},
  \bibinfo{pages}{083903} (\bibinfo{year}{2004}).

\bibitem[{\citenamefont{Nabekawa et~al.}(2005)\citenamefont{Nabekawa, Hasegawa,
  Takahashi, and Midorikawa}}]{Nabekawa}
\bibinfo{author}{\bibfnamefont{Y.}~\bibnamefont{Nabekawa}},
  \bibinfo{author}{\bibfnamefont{H.}~\bibnamefont{Hasegawa}},
  \bibinfo{author}{\bibfnamefont{E.~J.} \bibnamefont{Takahashi}},
  \bibnamefont{and}
  \bibinfo{author}{\bibfnamefont{K.}~\bibnamefont{Midorikawa}},
  \bibinfo{journal}{Phys. Rev. Lett.} \textbf{\bibinfo{volume}{94}},
  \bibinfo{pages}{043001} (\bibinfo{year}{2005}).

\bibitem[{\citenamefont{Hoshina et~al.}(2006)\citenamefont{Hoshina, Hishikawa,
  Kato, Sako, Yamanouchi, Takahashi, Nabekawa, and Midorikawa}}]{hoshina}
\bibinfo{author}{\bibfnamefont{K.}~\bibnamefont{Hoshina}},
  \bibinfo{author}{\bibfnamefont{A.}~\bibnamefont{Hishikawa}},
  \bibinfo{author}{\bibfnamefont{K.}~\bibnamefont{Kato}},
  \bibinfo{author}{\bibfnamefont{T.}~\bibnamefont{Sako}},
  \bibinfo{author}{\bibfnamefont{K.}~\bibnamefont{Yamanouchi}},
  \bibinfo{author}{\bibfnamefont{E.~J.} \bibnamefont{Takahashi}},
  \bibinfo{author}{\bibfnamefont{Y.}~\bibnamefont{Nabekawa}}, \bibnamefont{and}
  \bibinfo{author}{\bibfnamefont{K.}~\bibnamefont{Midorikawa}},
  \bibinfo{journal}{J. Phys. B} \textbf{\bibinfo{volume}{39}},
  \bibinfo{pages}{813} (\bibinfo{year}{2006}).

\bibitem[{\citenamefont{Ackermann}(2007)}]{Ackermann}
\bibinfo{author}{\bibfnamefont{W.}~\bibnamefont{Ackermann}},
  \bibinfo{journal}{Nat. Photon.} \textbf{\bibinfo{volume}{1}},
  \bibinfo{pages}{336} (\bibinfo{year}{2007}).

\bibitem[{\citenamefont{Shintake et~al.}(2008)}]{Shintake2008}
\bibinfo{author}{\bibfnamefont{T.}~\bibnamefont{Shintake}}
  \bibnamefont{et~al.}, \bibinfo{journal}{Nat. Photon.}
  \textbf{\bibinfo{volume}{2}}, \bibinfo{pages}{555} (\bibinfo{year}{2008}).

\bibitem[{\citenamefont{Sorokin
  et~al.}(2007{\natexlab{a}})\citenamefont{Sorokin, Bobashev, Feigl, Tiedtke,
  Wabnitz, and Richter}}]{Sorokin_PRL}
\bibinfo{author}{\bibfnamefont{A.~A.} \bibnamefont{Sorokin}},
  \bibinfo{author}{\bibfnamefont{S.~V.} \bibnamefont{Bobashev}},
  \bibinfo{author}{\bibfnamefont{T.}~\bibnamefont{Feigl}},
  \bibinfo{author}{\bibfnamefont{K.}~\bibnamefont{Tiedtke}},
  \bibinfo{author}{\bibfnamefont{H.}~\bibnamefont{Wabnitz}}, \bibnamefont{and}
  \bibinfo{author}{\bibfnamefont{M.}~\bibnamefont{Richter}},
  \bibinfo{journal}{Phys. Rev. Lett.} \textbf{\bibinfo{volume}{99}},
  \bibinfo{pages}{213002} (\bibinfo{year}{2007}{\natexlab{a}}).

\bibitem[{\citenamefont{Moshammer et~al.}(2007)\citenamefont{Moshammer, Jiang,
  Foucar, Rudenko, Ergler, Schr\"oter, L\"udemann, Zrost, Fischer, Titze
  et~al.}}]{Moshammer}
\bibinfo{author}{\bibfnamefont{R.}~\bibnamefont{Moshammer}},
  \bibinfo{author}{\bibfnamefont{Y.~H.} \bibnamefont{Jiang}},
  \bibinfo{author}{\bibfnamefont{L.}~\bibnamefont{Foucar}},
  \bibinfo{author}{\bibfnamefont{A.}~\bibnamefont{Rudenko}},
  \bibinfo{author}{\bibfnamefont{T.}~\bibnamefont{Ergler}},
  \bibinfo{author}{\bibfnamefont{C.~D.} \bibnamefont{Schr\"oter}},
  \bibinfo{author}{\bibfnamefont{S.}~\bibnamefont{L\"udemann}},
  \bibinfo{author}{\bibfnamefont{K.}~\bibnamefont{Zrost}},
  \bibinfo{author}{\bibfnamefont{D.}~\bibnamefont{Fischer}},
  \bibinfo{author}{\bibfnamefont{J.}~\bibnamefont{Titze}},
  \bibnamefont{et~al.}, \bibinfo{journal}{Phys. Rev. Lett.}
  \textbf{\bibinfo{volume}{98}}, \bibinfo{pages}{203001}
  (\bibinfo{year}{2007}).

\bibitem[{\citenamefont{Laarmann et~al.}(2005)}]{Laarmann}
\bibinfo{author}{\bibfnamefont{T.}~\bibnamefont{Laarmann}}
  \bibnamefont{et~al.}, \bibinfo{journal}{Phys. Rev. A}
  \textbf{\bibinfo{volume}{72}}, \bibinfo{pages}{023409}
  (\bibinfo{year}{2005}).

\bibitem[{\citenamefont{Wabnitz et~al.}(2002)}]{Wabnitz_Nature}
\bibinfo{author}{\bibfnamefont{H.}~\bibnamefont{Wabnitz}} \bibnamefont{et~al.},
  \bibinfo{journal}{Nature} \textbf{\bibinfo{volume}{420}},
  \bibinfo{pages}{482} (\bibinfo{year}{2002}).

\bibitem[{\citenamefont{Parker et~al.}(2006)}]{Parker_PRL_2006}
\bibinfo{author}{\bibfnamefont{J.~S.} \bibnamefont{Parker}}
  \bibnamefont{et~al.}, \bibinfo{journal}{Phys. Rev. Lett.}
  \textbf{\bibinfo{volume}{96}}, \bibinfo{pages}{133001}
  (\bibinfo{year}{2006}).

\bibitem[{\citenamefont{Parker et~al.}(2001)\citenamefont{Parker, Moore,
  Meharg, Dundas, and Taylor}}]{Parker2001}
\bibinfo{author}{\bibfnamefont{J.~S.} \bibnamefont{Parker}},
  \bibinfo{author}{\bibfnamefont{L.~R.} \bibnamefont{Moore}},
  \bibinfo{author}{\bibfnamefont{K.~J.} \bibnamefont{Meharg}},
  \bibinfo{author}{\bibfnamefont{D.}~\bibnamefont{Dundas}}, \bibnamefont{and}
  \bibinfo{author}{\bibfnamefont{K.~T.} \bibnamefont{Taylor}},
  \bibinfo{journal}{J. Phys. B} \textbf{\bibinfo{volume}{34}},
  \bibinfo{pages}{L69} (\bibinfo{year}{2001}).

\bibitem[{\citenamefont{Moore et~al.}(2009)}]{Moore}
\bibinfo{author}{\bibfnamefont{L.~R.} \bibnamefont{Moore}}
  \bibnamefont{et~al.}, \bibinfo{journal}{J. Phys. Conf. Ser.}
  \textbf{\bibinfo{volume}{194}}, \bibinfo{pages}{032055}
  (\bibinfo{year}{2009}).

\bibitem[{\citenamefont{Colgan and Pindzola}(2002)}]{Pindzola}
\bibinfo{author}{\bibfnamefont{J.}~\bibnamefont{Colgan}} \bibnamefont{and}
  \bibinfo{author}{\bibfnamefont{M.~S.} \bibnamefont{Pindzola}},
  \bibinfo{journal}{Phys. Rev. Lett.} \textbf{\bibinfo{volume}{88}},
  \bibinfo{pages}{173002} (\bibinfo{year}{2002}).

\bibitem[{\citenamefont{Feng and van~der Hart}(2003)}]{Feng2003}
\bibinfo{author}{\bibfnamefont{L.}~\bibnamefont{Feng}} \bibnamefont{and}
  \bibinfo{author}{\bibfnamefont{H.~W.} \bibnamefont{van~der Hart}},
  \bibinfo{journal}{J. Phys. B} \textbf{\bibinfo{volume}{36}},
  \bibinfo{pages}{L1} (\bibinfo{year}{2003}).

\bibitem[{\citenamefont{Laulan and Bachau}(2003)}]{Bachau2003}
\bibinfo{author}{\bibfnamefont{S.}~\bibnamefont{Laulan}} \bibnamefont{and}
  \bibinfo{author}{\bibfnamefont{H.}~\bibnamefont{Bachau}},
  \bibinfo{journal}{Phys. Rev. A} \textbf{\bibinfo{volume}{68}},
  \bibinfo{pages}{013409} (\bibinfo{year}{2003}).

\bibitem[{\citenamefont{Piraux et~al.}(2003)\citenamefont{Piraux, Bauer,
  Laulan, and Bachau}}]{Bachau_EPD}
\bibinfo{author}{\bibfnamefont{B.}~\bibnamefont{Piraux}},
  \bibinfo{author}{\bibfnamefont{J.}~\bibnamefont{Bauer}},
  \bibinfo{author}{\bibfnamefont{S.}~\bibnamefont{Laulan}}, \bibnamefont{and}
  \bibinfo{author}{\bibfnamefont{H.}~\bibnamefont{Bachau}},
  \bibinfo{journal}{Eur. Phys. J. D} \textbf{\bibinfo{volume}{26}},
  \bibinfo{pages}{7} (\bibinfo{year}{2003}).

\bibitem[{\citenamefont{Hu et~al.}(2005)\citenamefont{Hu, Colgan, and
  Collins}}]{Hu2005}
\bibinfo{author}{\bibfnamefont{S.~X.} \bibnamefont{Hu}},
  \bibinfo{author}{\bibfnamefont{J.}~\bibnamefont{Colgan}}, \bibnamefont{and}
  \bibinfo{author}{\bibfnamefont{L.~A.} \bibnamefont{Collins}},
  \bibinfo{journal}{J. Phys. B} \textbf{\bibinfo{volume}{38}},
  \bibinfo{pages}{L35} (\bibinfo{year}{2005}).

\bibitem[{\citenamefont{Foumouo et~al.}(2006)\citenamefont{Foumouo, Kamta,
  Edah, and Piraux}}]{Foumouo2006}
\bibinfo{author}{\bibfnamefont{E.}~\bibnamefont{Foumouo}},
  \bibinfo{author}{\bibfnamefont{G.~L.} \bibnamefont{Kamta}},
  \bibinfo{author}{\bibfnamefont{G.}~\bibnamefont{Edah}}, \bibnamefont{and}
  \bibinfo{author}{\bibfnamefont{B.}~\bibnamefont{Piraux}},
  \bibinfo{journal}{Phys. Rev. A} \textbf{\bibinfo{volume}{74}},
  \bibinfo{pages}{063409} (\bibinfo{year}{2006}).

\bibitem[{\citenamefont{Shakeshaft}(2007)}]{Shakeshaft2007}
\bibinfo{author}{\bibfnamefont{R.}~\bibnamefont{Shakeshaft}},
  \bibinfo{journal}{Phys. Rev. A} \textbf{\bibinfo{volume}{76}},
  \bibinfo{pages}{063405} (\bibinfo{year}{2007}).

\bibitem[{\citenamefont{Ivanov and Kheifets}(2007)}]{Ivanov2007}
\bibinfo{author}{\bibfnamefont{I.~A.} \bibnamefont{Ivanov}} \bibnamefont{and}
  \bibinfo{author}{\bibfnamefont{A.~S.} \bibnamefont{Kheifets}},
  \bibinfo{journal}{Phys. Rev. A} \textbf{\bibinfo{volume}{75}},
  \bibinfo{pages}{033411} (\bibinfo{year}{2007}).

\bibitem[{\citenamefont{Horner et~al.}(2007)\citenamefont{Horner, Morales,
  Rescigno, Mart\'{\i}n, and McCurdy}}]{Horner2007}
\bibinfo{author}{\bibfnamefont{D.~A.} \bibnamefont{Horner}},
  \bibinfo{author}{\bibfnamefont{F.}~\bibnamefont{Morales}},
  \bibinfo{author}{\bibfnamefont{T.~N.} \bibnamefont{Rescigno}},
  \bibinfo{author}{\bibfnamefont{F.}~\bibnamefont{Mart\'{\i}n}},
  \bibnamefont{and} \bibinfo{author}{\bibfnamefont{C.~W.}
  \bibnamefont{McCurdy}}, \bibinfo{journal}{Phys. Rev. A}
  \textbf{\bibinfo{volume}{76}}, \bibinfo{pages}{030701(R)}
  (\bibinfo{year}{2007}).

\bibitem[{\citenamefont{Nikolopoulos and
  Lambropoulos}(2007)}]{Nikolopoulos2007}
\bibinfo{author}{\bibfnamefont{L.~A.~A.} \bibnamefont{Nikolopoulos}}
  \bibnamefont{and}
  \bibinfo{author}{\bibfnamefont{P.}~\bibnamefont{Lambropoulos}},
  \bibinfo{journal}{J. Phys. B} \textbf{\bibinfo{volume}{40}},
  \bibinfo{pages}{1347} (\bibinfo{year}{2007}).

\bibitem[{\citenamefont{Feist et~al.}(2008)\citenamefont{Feist, Nagele,
  Pazourek, Persson, Schneider, Collins, and Burgd\"orfer}}]{Feist2008}
\bibinfo{author}{\bibfnamefont{J.}~\bibnamefont{Feist}},
  \bibinfo{author}{\bibfnamefont{S.}~\bibnamefont{Nagele}},
  \bibinfo{author}{\bibfnamefont{R.}~\bibnamefont{Pazourek}},
  \bibinfo{author}{\bibfnamefont{E.}~\bibnamefont{Persson}},
  \bibinfo{author}{\bibfnamefont{B.~I.} \bibnamefont{Schneider}},
  \bibinfo{author}{\bibfnamefont{L.~A.} \bibnamefont{Collins}},
  \bibnamefont{and}
  \bibinfo{author}{\bibfnamefont{J.}~\bibnamefont{Burgd\"orfer}},
  \bibinfo{journal}{Phys. Rev. A} \textbf{\bibinfo{volume}{77}},
  \bibinfo{pages}{043420} (\bibinfo{year}{2008}).

\bibitem[{\citenamefont{Guan et~al.}(2008)\citenamefont{Guan, Bartschat, and
  Schneider}}]{Guan2008}
\bibinfo{author}{\bibfnamefont{X.}~\bibnamefont{Guan}},
  \bibinfo{author}{\bibfnamefont{K.}~\bibnamefont{Bartschat}},
  \bibnamefont{and} \bibinfo{author}{\bibfnamefont{B.~I.}
  \bibnamefont{Schneider}}, \bibinfo{journal}{Phys. Rev. A}
  \textbf{\bibinfo{volume}{77}}, \bibinfo{pages}{043421}
  (\bibinfo{year}{2008}).

\bibitem[{\citenamefont{Foumouo et~al.}(2008)\citenamefont{Foumouo, Antoine,
  Piraux, Malegat, Bachau, and Shakeshaft}}]{Foumouo2008}
\bibinfo{author}{\bibfnamefont{E.}~\bibnamefont{Foumouo}},
  \bibinfo{author}{\bibfnamefont{P.}~\bibnamefont{Antoine}},
  \bibinfo{author}{\bibfnamefont{B.}~\bibnamefont{Piraux}},
  \bibinfo{author}{\bibfnamefont{L.}~\bibnamefont{Malegat}},
  \bibinfo{author}{\bibfnamefont{H.}~\bibnamefont{Bachau}}, \bibnamefont{and}
  \bibinfo{author}{\bibfnamefont{R.}~\bibnamefont{Shakeshaft}},
  \bibinfo{journal}{J. Phys. B} \textbf{\bibinfo{volume}{41}},
  \bibinfo{pages}{051001} (\bibinfo{year}{2008}).

\bibitem[{\citenamefont{Palacios et~al.}(2009)\citenamefont{Palacios, Rescigno,
  and McCurdy}}]{Palacios2009}
\bibinfo{author}{\bibfnamefont{A.}~\bibnamefont{Palacios}},
  \bibinfo{author}{\bibfnamefont{T.~N.} \bibnamefont{Rescigno}},
  \bibnamefont{and} \bibinfo{author}{\bibfnamefont{C.~W.}
  \bibnamefont{McCurdy}}, \bibinfo{journal}{Phys. Rev. A}
  \textbf{\bibinfo{volume}{79}}, \bibinfo{pages}{033402}
  (\bibinfo{year}{2009}).

\bibitem[{\citenamefont{Nepstad et~al.}(2010)\citenamefont{Nepstad, Birkeland,
  and F\o{}rre}}]{Nepstad2010}
\bibinfo{author}{\bibfnamefont{R.}~\bibnamefont{Nepstad}},
  \bibinfo{author}{\bibfnamefont{T.}~\bibnamefont{Birkeland}},
  \bibnamefont{and} \bibinfo{author}{\bibfnamefont{M.}~\bibnamefont{F\o{}rre}},
  \bibinfo{journal}{Phys. Rev. A} \textbf{\bibinfo{volume}{81}},
  \bibinfo{pages}{063402} (\bibinfo{year}{2010}).

\bibitem[{\citenamefont{Hasegawa et~al.}(2005)\citenamefont{Hasegawa,
  Takahashi, Nabekawa, Ishikawa, and Midorikawa}}]{Hasegawa_PRA_2005}
\bibinfo{author}{\bibfnamefont{H.}~\bibnamefont{Hasegawa}},
  \bibinfo{author}{\bibfnamefont{E.~J.} \bibnamefont{Takahashi}},
  \bibinfo{author}{\bibfnamefont{Y.}~\bibnamefont{Nabekawa}},
  \bibinfo{author}{\bibfnamefont{K.~L.} \bibnamefont{Ishikawa}},
  \bibnamefont{and}
  \bibinfo{author}{\bibfnamefont{K.}~\bibnamefont{Midorikawa}},
  \bibinfo{journal}{Phys. Rev. A} \textbf{\bibinfo{volume}{71}},
  \bibinfo{pages}{023407} (\bibinfo{year}{2005}).

\bibitem[{\citenamefont{Sorokin
  et~al.}(2007{\natexlab{b}})\citenamefont{Sorokin, Wellh\"ofer, Bobashev,
  Tiedtke, and Richter}}]{Sorokin2007}
\bibinfo{author}{\bibfnamefont{A.~A.} \bibnamefont{Sorokin}},
  \bibinfo{author}{\bibfnamefont{M.}~\bibnamefont{Wellh\"ofer}},
  \bibinfo{author}{\bibfnamefont{S.~V.} \bibnamefont{Bobashev}},
  \bibinfo{author}{\bibfnamefont{K.}~\bibnamefont{Tiedtke}}, \bibnamefont{and}
  \bibinfo{author}{\bibfnamefont{M.}~\bibnamefont{Richter}},
  \bibinfo{journal}{Phys. Rev. A} \textbf{\bibinfo{volume}{75}},
  \bibinfo{pages}{051402(R)} (\bibinfo{year}{2007}{\natexlab{b}}).

\bibitem[{\citenamefont{Rudenko et~al.}(2008)\citenamefont{Rudenko, Foucar,
  Kurka, Ergler, K\"uhnel, Jiang, Voitkiv, Najjari, Kheifets, L\"udemann
  et~al.}}]{Rudenko}
\bibinfo{author}{\bibfnamefont{A.}~\bibnamefont{Rudenko}},
  \bibinfo{author}{\bibfnamefont{L.}~\bibnamefont{Foucar}},
  \bibinfo{author}{\bibfnamefont{M.}~\bibnamefont{Kurka}},
  \bibinfo{author}{\bibfnamefont{T.}~\bibnamefont{Ergler}},
  \bibinfo{author}{\bibfnamefont{K.~U.} \bibnamefont{K\"uhnel}},
  \bibinfo{author}{\bibfnamefont{Y.~H.} \bibnamefont{Jiang}},
  \bibinfo{author}{\bibfnamefont{A.}~\bibnamefont{Voitkiv}},
  \bibinfo{author}{\bibfnamefont{B.}~\bibnamefont{Najjari}},
  \bibinfo{author}{\bibfnamefont{A.}~\bibnamefont{Kheifets}},
  \bibinfo{author}{\bibfnamefont{S.}~\bibnamefont{L\"udemann}},
  \bibnamefont{et~al.}, \bibinfo{journal}{Phys. Rev. Lett.}
  \textbf{\bibinfo{volume}{101}}, \bibinfo{pages}{073003}
  (\bibinfo{year}{2008}).

\bibitem[{\citenamefont{Briggs and Schmidt}(2000)}]{Briggs}
\bibinfo{author}{\bibfnamefont{J.~S.} \bibnamefont{Briggs}} \bibnamefont{and}
  \bibinfo{author}{\bibfnamefont{V.}~\bibnamefont{Schmidt}},
  \bibinfo{journal}{J. Phys. B} \textbf{\bibinfo{volume}{33}},
  \bibinfo{pages}{R1} (\bibinfo{year}{2000}).

\bibitem[{\citenamefont{Avaldi and Huetz}(2005)}]{Huetz}
\bibinfo{author}{\bibfnamefont{L.}~\bibnamefont{Avaldi}} \bibnamefont{and}
  \bibinfo{author}{\bibfnamefont{A.}~\bibnamefont{Huetz}}, \bibinfo{journal}{J.
  Phys. B} \textbf{\bibinfo{volume}{38}}, \bibinfo{pages}{S861}
  (\bibinfo{year}{2005}).

\bibitem[{\citenamefont{Samson et~al.}(1998)\citenamefont{Samson, Stolte, He,
  Cutler, Lu, and Bartlett}}]{Samson1998}
\bibinfo{author}{\bibfnamefont{J.~A.~R.} \bibnamefont{Samson}},
  \bibinfo{author}{\bibfnamefont{W.~C.} \bibnamefont{Stolte}},
  \bibinfo{author}{\bibfnamefont{Z.-X.} \bibnamefont{He}},
  \bibinfo{author}{\bibfnamefont{J.~N.} \bibnamefont{Cutler}},
  \bibinfo{author}{\bibfnamefont{Y.}~\bibnamefont{Lu}}, \bibnamefont{and}
  \bibinfo{author}{\bibfnamefont{R.~J.} \bibnamefont{Bartlett}},
  \bibinfo{journal}{Phys. Rev. A} \textbf{\bibinfo{volume}{57}},
  \bibinfo{pages}{1906} (\bibinfo{year}{1998}).

\bibitem[{\citenamefont{Selst{\o}¸ and Kvaal}(2010)}]{Selsto2010}
\bibinfo{author}{\bibfnamefont{S.}~\bibnamefont{Selst{\o}¸}} \bibnamefont{and}
  \bibinfo{author}{\bibfnamefont{S.}~\bibnamefont{Kvaal}}, \bibinfo{journal}{J.
  Phys. B} \textbf{\bibinfo{volume}{43}}, \bibinfo{pages}{065004}
  (\bibinfo{year}{2010}).

\bibitem[{\citenamefont{Kulander}(1987)}]{Kulander1987}
\bibinfo{author}{\bibfnamefont{K.~C.} \bibnamefont{Kulander}},
  \bibinfo{journal}{Phys. Rev. A} \textbf{\bibinfo{volume}{35}},
  \bibinfo{pages}{445} (\bibinfo{year}{1987}).

\bibitem[{\citenamefont{Fevens and Jiang}(1999)}]{Fevens1999}
\bibinfo{author}{\bibfnamefont{T.}~\bibnamefont{Fevens}} \bibnamefont{and}
  \bibinfo{author}{\bibfnamefont{H.}~\bibnamefont{Jiang}},
  \bibinfo{journal}{SIAM J. Sci. Comput.} \textbf{\bibinfo{volume}{21}},
  \bibinfo{pages}{255} (\bibinfo{year}{1999}).

\bibitem[{\citenamefont{Feuerstein and Thumm}(2003)}]{Feuerstein2003}
\bibinfo{author}{\bibfnamefont{B.}~\bibnamefont{Feuerstein}} \bibnamefont{and}
  \bibinfo{author}{\bibfnamefont{U.}~\bibnamefont{Thumm}}, \bibinfo{journal}{J.
  Phys. B} \textbf{\bibinfo{volume}{36}}, \bibinfo{pages}{707}
  (\bibinfo{year}{2003}).

\bibitem[{\citenamefont{Lein et~al.}(2000)\citenamefont{Lein, Gross, and
  Engel}}]{Lein2000}
\bibinfo{author}{\bibfnamefont{M.}~\bibnamefont{Lein}},
  \bibinfo{author}{\bibfnamefont{E.~K.~U.} \bibnamefont{Gross}},
  \bibnamefont{and} \bibinfo{author}{\bibfnamefont{V.}~\bibnamefont{Engel}},
  \bibinfo{journal}{Phys. Rev. Lett.} \textbf{\bibinfo{volume}{85}},
  \bibinfo{pages}{4707} (\bibinfo{year}{2000}).

\bibitem[{\citenamefont{Bauer and Ceccherini}(1999)}]{Bauer1999}
\bibinfo{author}{\bibfnamefont{D.}~\bibnamefont{Bauer}} \bibnamefont{and}
  \bibinfo{author}{\bibfnamefont{F.}~\bibnamefont{Ceccherini}},
  \bibinfo{journal}{Phys. Rev. A} \textbf{\bibinfo{volume}{60}},
  \bibinfo{pages}{2301} (\bibinfo{year}{1999}).

\bibitem[{\citenamefont{Lappas and van Leeuwen}(1998)}]{Lappas1998}
\bibinfo{author}{\bibfnamefont{D.~G.} \bibnamefont{Lappas}} \bibnamefont{and}
  \bibinfo{author}{\bibfnamefont{R.}~\bibnamefont{van Leeuwen}},
  \bibinfo{journal}{J. Phys. B} \textbf{\bibinfo{volume}{31}},
  \bibinfo{pages}{L249} (\bibinfo{year}{1998}).

\bibitem[{\citenamefont{Haan et~al.}(2002)\citenamefont{Haan, Wheeler, Panfili,
  and Eberly}}]{Haan2002}
\bibinfo{author}{\bibfnamefont{S.~L.} \bibnamefont{Haan}},
  \bibinfo{author}{\bibfnamefont{P.~S.} \bibnamefont{Wheeler}},
  \bibinfo{author}{\bibfnamefont{R.}~\bibnamefont{Panfili}}, \bibnamefont{and}
  \bibinfo{author}{\bibfnamefont{J.~H.} \bibnamefont{Eberly}},
  \bibinfo{journal}{Phys. Rev. A} \textbf{\bibinfo{volume}{66}},
  \bibinfo{pages}{061402} (\bibinfo{year}{2002}).

\bibitem[{\citenamefont{Zhang et~al.}(2010)\citenamefont{Zhang, Peng, Gong, and
  Morishita}}]{Zhang2010}
\bibinfo{author}{\bibfnamefont{Z.}~\bibnamefont{Zhang}},
  \bibinfo{author}{\bibfnamefont{L.-Y.} \bibnamefont{Peng}},
  \bibinfo{author}{\bibfnamefont{Q.}~\bibnamefont{Gong}}, \bibnamefont{and}
  \bibinfo{author}{\bibfnamefont{T.}~\bibnamefont{Morishita}},
  \bibinfo{journal}{Opt. Express} \textbf{\bibinfo{volume}{18}},
  \bibinfo{pages}{8976} (\bibinfo{year}{2010}).

\bibitem[{\citenamefont{Horner et~al.}(2008)\citenamefont{Horner, Rescigno, and
  McCurdy}}]{Horner2008}
\bibinfo{author}{\bibfnamefont{D.~A.} \bibnamefont{Horner}},
  \bibinfo{author}{\bibfnamefont{T.~N.} \bibnamefont{Rescigno}},
  \bibnamefont{and} \bibinfo{author}{\bibfnamefont{C.~W.}
  \bibnamefont{McCurdy}}, \bibinfo{journal}{Phys. Rev. A}
  \textbf{\bibinfo{volume}{77}}, \bibinfo{pages}{030703(R)}
  (\bibinfo{year}{2008}).

\bibitem[{\citenamefont{Lambropoulos et~al.}(2008)\citenamefont{Lambropoulos,
  Nikolopoulos, Makris, and Mihelic\ifmmode~\check{c}\else
  \v{c}\fi{}}}]{Lambropoulos_PRA_2008}
\bibinfo{author}{\bibfnamefont{P.}~\bibnamefont{Lambropoulos}},
  \bibinfo{author}{\bibfnamefont{L.~A.~A.} \bibnamefont{Nikolopoulos}},
  \bibinfo{author}{\bibfnamefont{M.~G.} \bibnamefont{Makris}},
  \bibnamefont{and}
  \bibinfo{author}{\bibfnamefont{A.}~\bibnamefont{Mihelic}}, \bibinfo{journal}{Phys. Rev. A} \textbf{\bibinfo{volume}{78}},
  \bibinfo{pages}{055402} (\bibinfo{year}{2008}).

\bibitem[{\citenamefont{Horner et~al.}(2010)\citenamefont{Horner, Rescigno, and
  McCurdy}}]{Horner_2010}
\bibinfo{author}{\bibfnamefont{D.~A.} \bibnamefont{Horner}},
  \bibinfo{author}{\bibfnamefont{T.~N.} \bibnamefont{Rescigno}},
  \bibnamefont{and} \bibinfo{author}{\bibfnamefont{C.~W.}
  \bibnamefont{McCurdy}}, \bibinfo{journal}{Phys. Rev. A}
  \textbf{\bibinfo{volume}{81}}, \bibinfo{pages}{023410}
  (\bibinfo{year}{2010}).

\bibitem[{\citenamefont{Piraux et~al.}(2008)\citenamefont{Piraux, Foumouo,
  Antoine, and Bachau}}]{Piraux2008}
\bibinfo{author}{\bibfnamefont{B.}~\bibnamefont{Piraux}},
  \bibinfo{author}{\bibfnamefont{E.}~\bibnamefont{Foumouo}},
  \bibinfo{author}{\bibfnamefont{P.}~\bibnamefont{Antoine}}, \bibnamefont{and}
  \bibinfo{author}{\bibfnamefont{H.}~\bibnamefont{Bachau}},
  \bibinfo{journal}{J. Phys. Conf. Ser.} \textbf{\bibinfo{volume}{141}},
  \bibinfo{pages}{012013} (\bibinfo{year}{2008}).

\bibitem[{\citenamefont{Birkeland et~al.}(2010)\citenamefont{Birkeland,
  Nepstad, and F\o{}rre}}]{Birkeland2010}
\bibinfo{author}{\bibfnamefont{T.}~\bibnamefont{Birkeland}},
  \bibinfo{author}{\bibfnamefont{R.}~\bibnamefont{Nepstad}}, \bibnamefont{and}
  \bibinfo{author}{\bibfnamefont{M.}~\bibnamefont{F\o{}rre}},
  \bibinfo{journal}{Phys. Rev. Lett.} \textbf{\bibinfo{volume}{104}},
  \bibinfo{pages}{163002} (\bibinfo{year}{2010}).

\bibitem[{\citenamefont{Wilson and Lindsay}(1935)}]{Wilson1935}
\bibinfo{author}{\bibfnamefont{W.~S.} \bibnamefont{Wilson}} \bibnamefont{and}
  \bibinfo{author}{\bibfnamefont{R.~B.} \bibnamefont{Lindsay}},
  \bibinfo{journal}{Phys. Rev.} \textbf{\bibinfo{volume}{47}},
  \bibinfo{pages}{681} (\bibinfo{year}{1935}).

\bibitem[{\citenamefont{Gorini et~al.}(1976)\citenamefont{Gorini, Kossakowski,
  and Sudarshan}}]{Gorini1976}
\bibinfo{author}{\bibfnamefont{V.}~\bibnamefont{Gorini}},
  \bibinfo{author}{\bibfnamefont{A.}~\bibnamefont{Kossakowski}},
  \bibnamefont{and}
  \bibinfo{author}{\bibfnamefont{E.}~\bibnamefont{Sudarshan}},
  \bibinfo{journal}{J. Math. Phys.} \textbf{\bibinfo{volume}{17}},
  \bibinfo{pages}{821} (\bibinfo{year}{1976}).

\bibitem[{\citenamefont{Lindblad}(1976)}]{Lindblad1976}
\bibinfo{author}{\bibfnamefont{G.}~\bibnamefont{Lindblad}},
  \bibinfo{journal}{Comm. Math. Phys.} \textbf{\bibinfo{volume}{48}},
  \bibinfo{pages}{119} (\bibinfo{year}{1976}).

\bibitem[{\citenamefont{Madsen et~al.}(2000)\citenamefont{Madsen, Nikolopoulos,
  and Lambropoulos}}]{Madsen2000}
\bibinfo{author}{\bibfnamefont{L.~B.} \bibnamefont{Madsen}},
  \bibinfo{author}{\bibfnamefont{L.~A.~A.} \bibnamefont{Nikolopoulos}},
  \bibnamefont{and}
  \bibinfo{author}{\bibfnamefont{P.}~\bibnamefont{Lambropoulos}},
  \bibinfo{journal}{Eur. Phys. J. D} \textbf{\bibinfo{volume}{10}},
  \bibinfo{pages}{67} (\bibinfo{year}{2000}).

\bibitem[{\citenamefont{Smyth et~al.}(1998)\citenamefont{Smyth, Parker, and
  Taylor}}]{Smyth1981}
\bibinfo{author}{\bibfnamefont{E.~S.} \bibnamefont{Smyth}},
  \bibinfo{author}{\bibfnamefont{J.~S.} \bibnamefont{Parker}},
  \bibnamefont{and} \bibinfo{author}{\bibfnamefont{K.}~\bibnamefont{Taylor}},
  \bibinfo{journal}{Comput. Phys. Commun.} \textbf{\bibinfo{volume}{114}},
  \bibinfo{pages}{1} (\bibinfo{year}{1998}).

\bibitem[{\citenamefont{Leth et~al.}(2009)\citenamefont{Leth, Madsen, and
  M\o{}lmer}}]{Leth2009}
\bibinfo{author}{\bibfnamefont{H.~A.} \bibnamefont{Leth}},
  \bibinfo{author}{\bibfnamefont{L.~B.} \bibnamefont{Madsen}},
  \bibnamefont{and}
  \bibinfo{author}{\bibfnamefont{K.}~\bibnamefont{M\o{}lmer}},
  \bibinfo{journal}{Phys. Rev. Lett.} \textbf{\bibinfo{volume}{103}},
  \bibinfo{pages}{183601} (\bibinfo{year}{2009}).

\bibitem[{\citenamefont{Feit et~al.}(1982)\citenamefont{Feit, Fleck, and
  Steiger}}]{Feit1982}
\bibinfo{author}{\bibfnamefont{M.}~\bibnamefont{Feit}},
  \bibinfo{author}{\bibfnamefont{J.}~\bibnamefont{Fleck}}, \bibnamefont{and}
  \bibinfo{author}{\bibfnamefont{A.}~\bibnamefont{Steiger}},
  \bibinfo{journal}{J. Comput. Phys.} \textbf{\bibinfo{volume}{47}},
  \bibinfo{pages}{412} (\bibinfo{year}{1982}).

\bibitem[{\citenamefont{Birkeland and Nepstad}()}]{pyprop}
\bibinfo{author}{\bibfnamefont{T.}~\bibnamefont{Birkeland}} \bibnamefont{and}
  \bibinfo{author}{\bibfnamefont{R.}~\bibnamefont{Nepstad}},
  \emph{\bibinfo{title}{Pyprop}},
  \bibinfo{howpublished}{\url{http://pyprop.googlecode.com}}.

\bibitem[{\citenamefont{Manolopoulos}(2002)}]{Manolopoulos2002}
\bibinfo{author}{\bibfnamefont{D.~E.} \bibnamefont{Manolopoulos}},
  \bibinfo{journal}{J. Chem. Phys.} \textbf{\bibinfo{volume}{117}},
  \bibinfo{pages}{9552} (\bibinfo{year}{2002}).

\end{thebibliography}
\end{document}